\documentclass[pra,twocolumn]{revtex4-1}

\usepackage{amsfonts,amssymb,amsmath}
\usepackage[]{graphics,graphicx,epsfig}
\usepackage{amsthm}
\usepackage{color}

\def\identity{\leavevmode\hbox{\small1\kern-3.8pt\normalsize1}}

\newtheorem{propo}{Proposition}
\newcommand{\be}{\begin{eqnarray} \begin{aligned}}
\newcommand{\ee}{\end{aligned} \end{eqnarray} }
\newcommand{\bpr}{\begin{propo}}
\newcommand{\epr}{\end{propo}}
\newcommand{\bpf}{\begin{proof}}
\newcommand{\epf}{\end{proof}}
\newcommand{\ket}[1]{\left | #1 \right\rangle}

\newcommand{\Tr}{\mathrm{Tr}}

\renewcommand{\epsilon}{\varepsilon}

\usepackage[T1]{fontenc}

\begin{document}

\title{Genuinely multi-point temporal quantum correlations \\ and universal measurement-based quantum computing}

\author{Marcin Markiewicz$^{1}$, Anna Przysi\k{e}\.zna$^{1}$, Stephen Brierley$^{2}$ and Tomasz Paterek$^{3,4}$}
\affiliation{
$^1$Institute of Theoretical Physics and Astrophysics, University of Gda\'nsk, 80-952 Gda\'nsk, Poland\\ 
$^2$Heilbronn Institute for Mathematical Research, Department of Mathematics,
University of Bristol, Bristol BS8 1TW, UK\\ 
$^3$School of Physical and Mathematical Sciences, Nanyang Technological University, Singapore\\
$^4$Centre for Quantum Technologies, National University of Singapore, Singapore}

\begin{abstract}
We introduce a constructive procedure that maps all spatial correlations of a broad class of states into temporal correlations between general quantum measurements. 
This allows us to present temporal phenomena analogous to genuinely multipartite nonlocal phenomena, such as Greenberger-Horne-Zeilinger correlations, which do not exist if only projective measurements on qubits are considered.
The map is applied to certain lattice systems in order to replace one spatial dimension with a temporal one, without affecting measured correlations. 
We use this map to show how repeated application of a 1d-cluster-gate leads to universal one-way quantum computing when supplemented with the general measurements. 
\end{abstract}

\maketitle

\emph{Introduction.} Quantum mechanics treats space and time very differently.
Whereas spatial coordinates are represented by operators, usually time enters as a number parameterising sequences of events.
One might expect that quantum predictions in a purely spatial domain are very different from those in a purely temporal domain.
To the contrary, here we show that for a broad class of states quantum expectation values measured on several particles in different spatial locations can be mapped to those measured on a single system at different instances of time.

Temporal correlations have been widely studied, beginning with the seminal work of Leggett and Garg who considered them in the context of macroscopic realism~\cite{LG85,LG_REVIEW}.
Brukner \emph{et al.} rephrased the scenario to include more observables at a single instant of time and showed a task that is solved more efficiently by non-macrorealistic system~\cite{BTCV04}.
Both approaches were further generalised to involve many instances of time~\cite{B09, AHW10, MKTLSPK13,Z10}
and a semidefinite program is known which decides whether a probability distribution can be realised in a sequential manner~\cite{BMKG13}.
All these works considered projective measurements and show that temporal measurement statistics are indeed different from the spatial ones:
for qubits there is no temporal analog of genuinely multipartite entanglement~\cite{DCT99}, no monogamy of entanglement~\cite{CCW00}, and temporal Tsirelson bounds can be higher~\cite{TSI}.

Here we show that some of these differences disappear as soon as one considers general (POVM) measurements in place of the projective ones.
Our approach is based on the fact that every pure quantum state of finite dimension can be expressed in the form of a Matrix Product State (MPS) and therefore, under some constraints on the entanglement structure, can be generated in a sequential manner \cite{Schon}. We utilize the sequential procedure to construct a series of POVM measurements on a single particle giving rise to temporal correlations identical to the spatial correlations of the corresponding MPS, even if the latter is genuinely multiparty entangled.
Moreover, we generalize the procedure to a broad class of quantum states, that cannot be generated sequentially in one dimension but can be represented as higher dimensional quantum lattice systems.  Spatial correlations in such systems are mapped into spatio-temporal ones reducing the lattice dimension by one.

As an important example we consider two dimensional (2D) cluster states, which are a universal resource for one-way quantum computing \cite{RB01, RBB03}. 
The spatial process of one-way computing on a 2D cluster state is transformed into a temporal process involving the sequential application of a gate preparing only 1D-cluster states.
The universality of this case is shown to follow from the use of POVMs; allowing only projective measurements results in correlations that can be simulated classically.


\emph{Temporal quantum correlations.} The  temporal correlation function for a sequence of $N$  generalized  quantum measurements is defined as follows:
\begin{equation*}
E_{m_1,\dots, m_N} = \! \! \! \! \! \sum_{i_1,\dots,i_N} \! \! \! i_1 \dots i_N P(i_1,\dots,i_N|m_1,\dots,m_N),
\end{equation*}
where
\begin{eqnarray}
\label{GenCondProb}
&&P(i_1,\dots,i_N|m_1,\dots,m_N) = P(i_1| m_1) P(i_2 | i_1,m_1,m_2)\times\nonumber\\
&&\dots P(i_N | i_1,\dots,i_{N-1},m_1,\dots,m_{N}),
\end{eqnarray}
is the probability to observe a particular sequence of outcomes $\{i_1,\dots,i_N\}$ conditioned on the settings $\{m_1,\dots,m_N\}$.

Let us first briefly discuss the case of sequential projective measurements on a qubit prepared in an initial state described by a Bloch vector $\vec s$.
The qubit is measured at time instances $t_1,\dots, t_N$ with the corresponding dichotomic observables parameterised by Bloch vectors $\vec m_1, \dots, \vec m_N$.
Sequential projective measurements on a single qubit form a Markov chain \cite{BTCV04,Z10}, so that the conditional probabilities in \eqref{GenCondProb} fulfil the Markov property:
\begin{equation}
P(i_k|i_1,\dots,i_{k-1},\vec m_1,\dots,\vec m_{k})=P(i_k|i_{k-1},\vec m_{k-1},\vec m_{k}).
\end{equation}
Straightforward calculation of the temporal correlations reveals that for odd and even $N$ we have respectively:
\begin{eqnarray}
E_{\vec m_1, \dots, \vec m_N} &=&(\vec m_1 \cdot \vec s)(\vec m_2 \cdot \vec m_3) \dots (\vec m_{N-1} \cdot \vec m_{N}),\nonumber\\
E_{\vec m_1, \dots, \vec m_N} &=&(\vec m_1 \cdot \vec m_2)(\vec m_3 \cdot \vec m_4) \dots (\vec m_{N-1} \cdot \vec m_{N}), \label{QUBIT_DECORR}
\end{eqnarray}
demonstrating that temporal correlations between outcomes of multiple projective measurements on a qubit always factorize into correlations between at most two measurements.
In this sense a qubit never gives rise to genuine multi-point correlations in time.

For generalized measurements, defined by measurement operators $\{M_k\}$, the  probabilities in \eqref{GenCondProb} read:
\begin{equation}
P(i_k|i_1,\dots,i_{k-1},m_1,\dots,m_{k})=\Tr \left(\rho_k M_k^{\dagger} M_k \right),
\end{equation}
where the post-measurement state is defined by a recursive formula 
\begin{equation}
\label{postmstate}
\rho_{k+1}=\left(M_{k} \rho_{k} M_{k}^{\dagger}\right)/\Tr\left( M_{k} \rho_{k} M_{k}^{\dagger}\right).
\end{equation}
The correlations of a sequence of POVM measurements depend directly on the measurement operators $M_k$ via the post-measurement states \eqref{postmstate}.
Different sets of $\{M_k\}$ correspond to different possible physical implementations of given POVM elements $E_k=M_k^{\dagger}M_k$. 
This is in contrast to the spatially-separated scenario where all necessary information is given by $E_k$.

\begin{figure*}
\includegraphics[width=0.9\textwidth]{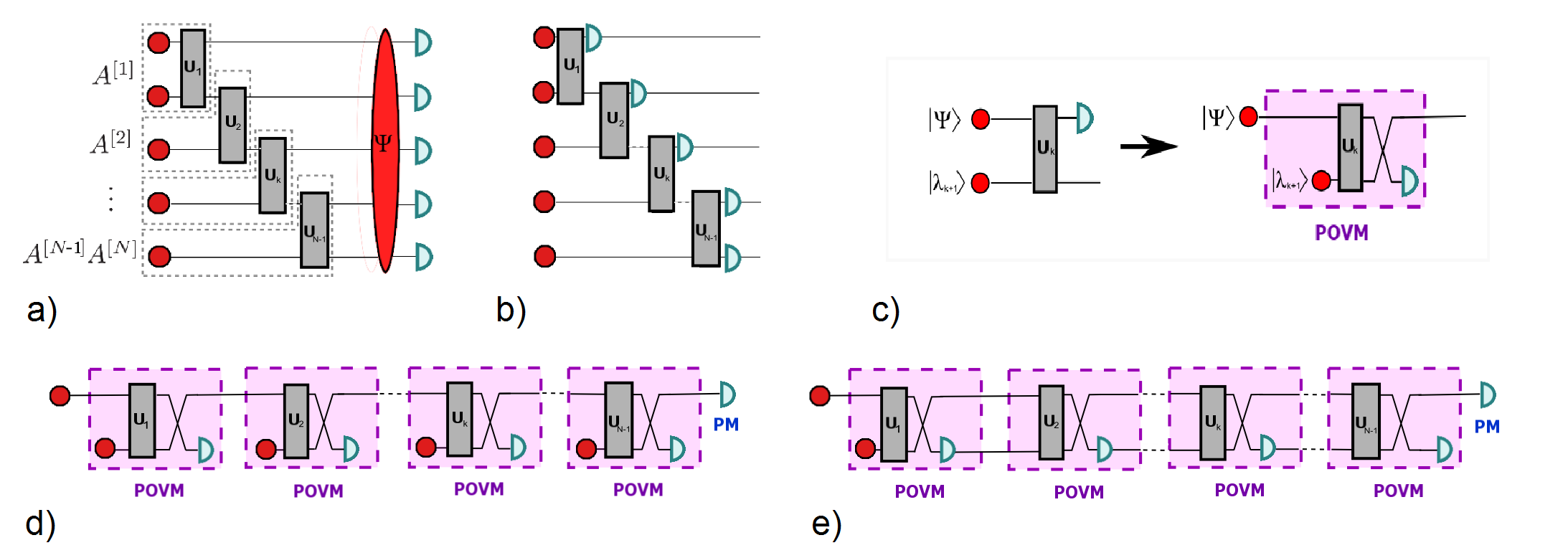}
\caption{Sequential generation of a multipartite MPS and its temporal counterpart. a) Two-particle unitary gates $U_k$ (grey rectangles) are sequentially performed on particles (red circles) giving the state $\Psi$ (big red shape). In the end, projective measurements (blue shapes) are conducted on the particles. Dashed lines indicate how unitary gates are related to MPS matrices of Eq.~(\ref{MPS}). b) Measurements can be shifted in time. c) Every gate followed by a measurement can equivalently be represented as a POVM. d) The circuit can be rearranged into a sequence of quantum channels performing POVM measurements. e) We can reduce the required resources to two particles. After each measurement, one of them is reset and recycled into the remaining protocol.}
\label{mpsgen}
\end{figure*}

\emph{Sequential generation of quantum states}. 
A vital part of our map between spatial and temporal correlations relies on the knowledge of sequential generation of a quantum state.
We say that a state can be generated sequentially, if it can be prepared from a product state by a sequential application of unitary operations on blocks of parties of a given size. Let us first consider a one dimensional lattice system, in which at each node there is a $d$-level quantum particle.
It was demonstrated by Sch\"on \emph{et al.} \cite{Schon} that sequentially generated states can be written as Matrix Product States (MPS) and, conversely, any MPS can be generated sequentially.

The MPS representation~\cite{Klumper2, Klumper3, Fannes, DMRG} is an efficient method of describing multipartite quantum states, most often used in the context of one-dimensional spin systems with local interactions. Here we use open boundary conditions for which the MPS form of a state is given by:
\begin{equation}
\label{MPS}
\ket{\Psi}=\sum_{i_1...i_n = 1}^d A^{[1]}_{i_1}...A_{i_N}^{[N]} \ket{i_1...i_N}.
\end{equation}
The first and the last matrices are vectors and each matrix $A_{i_n}^{[n]}$ has a maximum dimension $D \times D$. The parameter $D$, called the bond dimension, is the largest rank of the reduced density matrix with respect to every cut.  
It was shown in Ref.~\cite{Vidal2003} that the rank of the reduced density matrix is a measure of entanglement and therefore $D$ contains information about the entanglement structure of the state.

To sequentially generate an MPS of a bond dimension $D$ we use the method proposed in Ref.~\cite{BGWVC08}. Starting from a chain of $N$ initially uncorrelated $d$-level particles we apply unitary operations on $m$ neighbouring particles in a sequential manner. As a result we obtain a state with bond dimension at most $D=d^{m-1}$. Given the MPS form of a state~\eqref{MPS}, the required unitaries can be computed using the singular value decomposition \cite{Schon}. 
Within this paper we restrict our attention to the case of bipartite unitaries, \emph{i.e.} $m=2$, which is depicted in Fig.~\ref{mpsgen}a.


\emph{Mapping from spatial to temporal correlations}. 
As described, any MPS of qu$d$its with bond dimension $D\leq d$ can be generated by a sequential application of bipartite unitary gates. 
Now we utilize this scheme to find a sequence of measurements performed on a \emph{single} particle such that the correlations between outcomes of these measurements  are exactly the same as spatial correlations measured on the MPS. 
Note the crucial requirement that a single particle evolving in time is measured.
If higher-dimensional systems are allowed the task becomes simpler and trivialises for systems with dimension equal to the total dimension of the MPS,
as then one can simply measure in time incomplete projective measurements given by the local measurements on the MPS.

The transition from the spatial to the temporal domain is depicted in Fig.~$1$. 
At the $k$-th step of the preparation procedure one of the particles prepared in the previous step interacts via the gate $U_k$ with the particle that has not been used up to now.
The latter particle is next subject to the gate $U_{k+1}$, whereas the former particle is left untouched during the rest of the procedure. 
This important characteristic of the sequential preparation allows one to perform a projective measurement on the first particle right after the gate $U_k$ is applied without disturbing the later process of preparing the state (see Fig. \ref{mpsgen}b).
The entire protocol can equivalently be seen as a sequence of quantum channels (Fig. \ref{mpsgen}c) with a single particle input and output that realise $d$-outcome POVM measurements (Fig. \ref{mpsgen}d). 
Let us parametrize by $\alpha_k$ the rank one projective measurement on the $k$-th particle.
The corresponding measurement operators $\{M^{(k)}\}$ can be determined from the equation~\cite{Book.Nielsen.Chuang}:
\begin{eqnarray}
\label{povm1}
S(U_k(\ket{\psi}\otimes\ket{\lambda_{k+1}}))&=&\sum_{i=1}^d\left(M^{(k)}_i\ket{\psi}\right)\otimes V(\alpha_k)\ket{i},
\end{eqnarray}
where $S$ is a swap operator, $V(\alpha_k)$ is a unitary rotation from the standard basis to the measurement basis of the $k$-th observer, $M^{(k)}_i$ is the  measurement operator corresponding to the outcome $i$, $|\lambda_{k+1}\rangle$ is the initial state of the $(k+1)$-st particle before the preparation takes place, and $\ket{\psi}$ is an arbitrary state. Note that in the presented mapping, the $(k+1)$-st particle of the MPS is mapped into an ancilla of the $k$-th POVM, the $k$-th projective measurement is mapped into the $k$-th POVM for $k=1,\dots,N-1$ and the last projective measurement is mapped into itself (Fig. \ref{mpsgen}d).

Finally, if we recycle the qu$d$its, we can implement  the construction with only two of them (Fig. \ref{mpsgen}e). 
A similar process of recycling of qubits has already been used in an experimental realization of Shor's algorithm \cite{MLAZO12} in the circuit model of quantum  computation.
Summing up, due to equivalence of quantum circuits depicted in Fig.~$1$, all quantum predictions of arbitrary pure MPS of $N$ qu$d$its, with bond dimension $D\leq d$, can be reconstructed by temporal consecutive $d$-outcome POVM measurements followed by a projective measurement in the final step.


\emph{Genuinely multi-point temporal correlations}. 
In the context of quantum information and quantum foundations many important multi-qubit states such as GHZ~\cite{GHZ90}, W \cite{W} and 1D cluster~\cite{BR01} states, have bond dimension $D=2$, hence they can be generated sequentially with bipartite unitaries.
To create GHZ state we apply C-not gates on a chain of qubits  prepared in a state $\ket{+}\ket{0}\dots\ket{0}$, whereas 
in the case of a 1D cluster we use C-phase gates on a state $\ket{+}\dots\ket{+}$. Both of these states are genuinely $N$-partite entangled, however their entanglement properties are very different \cite{1dClusterEnt}.

We show explicitly a set of measurements producing multi-point temporal correlations.
Consider local projective measurements performed on GHZ and cluster states.
Denote the angles (on the Bloch sphere) parametrizing the measurement on the $k$-th particle by $\{\phi_k,\theta_k\}$.
Solving equation (\ref{povm1}) it is easy to obtain the corresponding $k$-th POVM measurement operators entering the temporal scenario.
In the case of a GHZ state we find:
\begin{eqnarray}
\label{povmghz}
M^{(k)}_{-1}&=&\left( \begin{array}{cc}
e^{i\phi_k}\sin(\theta_k/2) & 0  \\
0 & -\cos(\theta_k/2)  \end{array} \right),\nonumber\\
M^{(k)}_{1}&=&\left( \begin{array}{cc}
\cos(\theta_k/2) & 0  \\
0 & e^{-i\phi_k}\sin(\theta_k/2)  \end{array} \right),
\end{eqnarray}
whereas for a cluster state they read:
\begin{eqnarray}
\label{povm1dcl}
M^{(k)}_{-1}&=&\frac{1}{\sqrt{2}}\left( \begin{array}{cc}
e^{i\phi_k}\sin(\theta_k/2) & -\cos(\theta_k/2)  \\
e^{i\phi_k}\sin(\theta_k/2) & \cos(\theta_k/2)  \end{array} \right),\nonumber\\
M^{(k)}_{1}&=&\frac{1}{\sqrt{2}}\left( \begin{array}{cc}
\cos(\theta_k/2) & e^{-i\phi_k}\sin(\theta_k/2)  \\
\cos(\theta_k/2) & -e^{-i\phi_k}\sin(\theta_k/2)  \end{array} \right).
\end{eqnarray}
Interestingly, both of them give rise to the same POVM elements:
\begin{eqnarray}
E^{(k)}_{-1}&=&\left( \begin{array}{cc}
\sin^2(\theta_k/2) & 0  \\
0 & \cos^2(\theta_k/2)  \end{array} \right),\nonumber\\
E^{(k)}_{1}&=&\left( \begin{array}{cc}
\cos^2(\theta_k/2) & 0  \\
0 & \sin^2(\theta_k/2)  \end{array} \right).
\label{ghzE}
\end{eqnarray}

The mapping from spatial to temporal correlations  allows us to obtain the correlation functions for local measurements on both the GHZ and 1D cluster states by performing sequential POVM measurements \eqref{povmghz} and \eqref{povm1dcl}  on a single qubit followed by a projective one at the last step. 
Note that in the above construction, the POVM elements (\ref{ghzE}) do not depend on the phase $\phi$, whereas the measurement operators \eqref{povmghz} and \eqref{povm1dcl}  directly depend on this parameter. This shows that the information about the phase is solely encoded in the way the state collapses at each stage of the sequence of measurements, and illustrates that knowledge about POVM elements alone is not sufficient to determine temporal correlations.


\emph{Mapping from spatial to spatio-temporal correlations}.
In case the bond dimension exceeds the dimension of a single quantum system under consideration, multipartite unitaries are necessary for the sequential generation, whereas our mapping works for bipartite unitaries only. This difficulty can be overcome to some extent by arranging single systems into an $r$-dimensional hypercubic lattice with a $d$-level quantum system at each node, such that sequential generation is possible along at least one spatial dimension.
This dimension can be mapped to time as follows.
We first generate the quantum state of the particles placed in the $(r-1)$-dimensional slice perpendicular to the distinguished dimension. 
Now, all the \emph{consecutive} measurements along the distinguished dimension are mapped into single-particle POVMs in complete analogy to the 1D case.
The details  are presented in Appendix A.

\emph{Measurement-based quantum computing.} Universal one-way quantum computing is a processing of quantum information based on single-qubit projective measurements performed on a resource state such as a 2D cluster state \cite{BR01}.
Each elementary gate from the circuit model can be implemented by a sequence of such measurements, although the sequence may depend on previous measurement outcomes. What is important, the classical information about outcomes has to be sent in a single direction \cite{RBB03}. Therefore,  we can adapt our procedure of transforming spatial correlations of a 2D cluster state into spatio-temporal ones, where the temporal direction is defined by the direction of classical information transfer. 
In effect, the computing can be performed by repeatedly preparing a 1D cluster state followed by local POVMs.

The idea of cluster state recycling has already appeared in the works of Raussendorf and Briegel \cite{BR01, RBB03} and  in, for example, proposals for cluster states built from optical lattices \cite{Devitt+09}. 
The entire 2D cluster state is not produced at the start of the computation but rather prepared just in time. 
Our scheme is similar in spirit, but only requires a 1D array of qubits in comparison to the 2D sub-cluster required to implement entangling gates via measurements.

\begin{figure}
\includegraphics[width=0.45\textwidth]{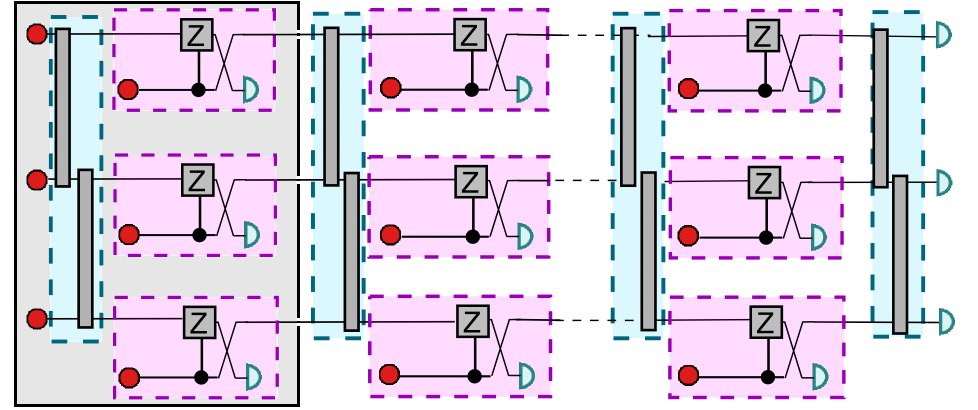}
\caption{One-way quantum computing in time. In a standard spatial implementation measurements are performed in a sequence along one dimension of a 2D cluster. This can be mapped into spatio-temporal scenario as follows.
First a vertical 1D cluster state (a blue-shaded rectangle) is prepared in the usual way.
Next, these qubits undergo local POVM measurements, which effectively entangle second column of qubits with the prepared 1D cluster, and measure the qubits of the 1D cluster (that would never again take part in the computation, and therefore can be recycled). 
Since the contents of the grey rectangle are repeated throughout the computation, universal one-way quantum computing requires a single `gate' preparing 1D cluster state that after each preparation is followed by generalized measurements.
In the last stage local projective measurements need to be performed.
}
\label{oneway_fig}
\end{figure}

The universality of the scheme given in Fig.~\ref{oneway_fig} relies on the use of POVMs.
To see this, first note that any computation arising by directly measuring a 1D cluster state, either with projective or POVM measurements, can be efficiently simulated classically~\cite{N05}.
Now consider a single 1D-cluster-gate, i.e. a sequence of c-phase gates preparing a 1D cluster state, and allow for projective or POVM measurements after each usage of this gate plus feed forward of classical information.
The scheme with projective measurements can still be efficiently simulated classically because these measurements uncorrelate future results from the previous ones, see Eq.~(\ref{QUBIT_DECORR}), and therefore one can repeatedly use the simulation of Ref.~\cite{N05}.
POVMs do not destroy the temporal correlations and as Fig.~\ref{oneway_fig} shows allow for universal computation.

A very simple example of entangling gates plus POVMs corresponds to the circuit model. In the trivial case of a single POVM element that is simply a one-qubit unitary rotation $E_1 = U$ we clearly have universal quantum computation. 
Our results demonstrate that non-trivial POVMs, where particles are really measured, can be used to promote the 1D lattice of qubits to universal quantum computation.
These non-trivial POVMs give rise to temporal quantum correlations that replace part of spatial correlations arising from 2D cluster states.

\emph{Conclusions.} 
We obtain genuinely multi-point temporal quantum correlations in a scenario of sequential generalized quantum measurements on a single particle. In this way we recover in a temporal experiment the statistics of results arising from local projective measurements of any multipartite quantum state of qu$d$its that has MPS representation with bond dimension $D\leq d$. Our approach can also be applied to a broad class of generalized graph states, subject to the restriction that they can be generated in a sequential way with respect to one spatial dimension. Spatial correlations of such graph states are then mapped into spatio-temporal ones reducing the space dimension by one. 

We show that repeatedly preparing a 1D cluster state followed by (non-trivial) local POVMs is universal for quantum computing. Our model allows for a resource reduction from $N^2$ qubits of a 2D cluster to $2N$ qubits of a 1D cluster with local POVMs, whereas the total number of entangling operations used for the computation is the same in both architectures. Our construction proves that genuinely multi-point temporal quantum correlations are a resource for quantum computing.

\emph{Acknowledgements}. We would like to thank Dan Browne, Alastair Kay, Adrian Kosowski, Marcin Paw{\l}owski,  Marek \.Zukowski, and anonymous referees for helpful remarks. MM and AP are supported by the International PhD Project ``Physics of future quantum-based information technologies''
grant MPD/2009-3/4 from Foundation for Polish Science and  by the University of Gda\'nsk grant 538-5400-B169-13.
This work is supported by the National Research Foundation, the Ministry of Education of Singapore grant no. RG98/13, start-up grant of the Nanyang Technological University, and NCN Grant No. 2012/05/E/ST2/02352.

\appendix
\section{}

\begin{figure*}
\includegraphics[width=\textwidth]{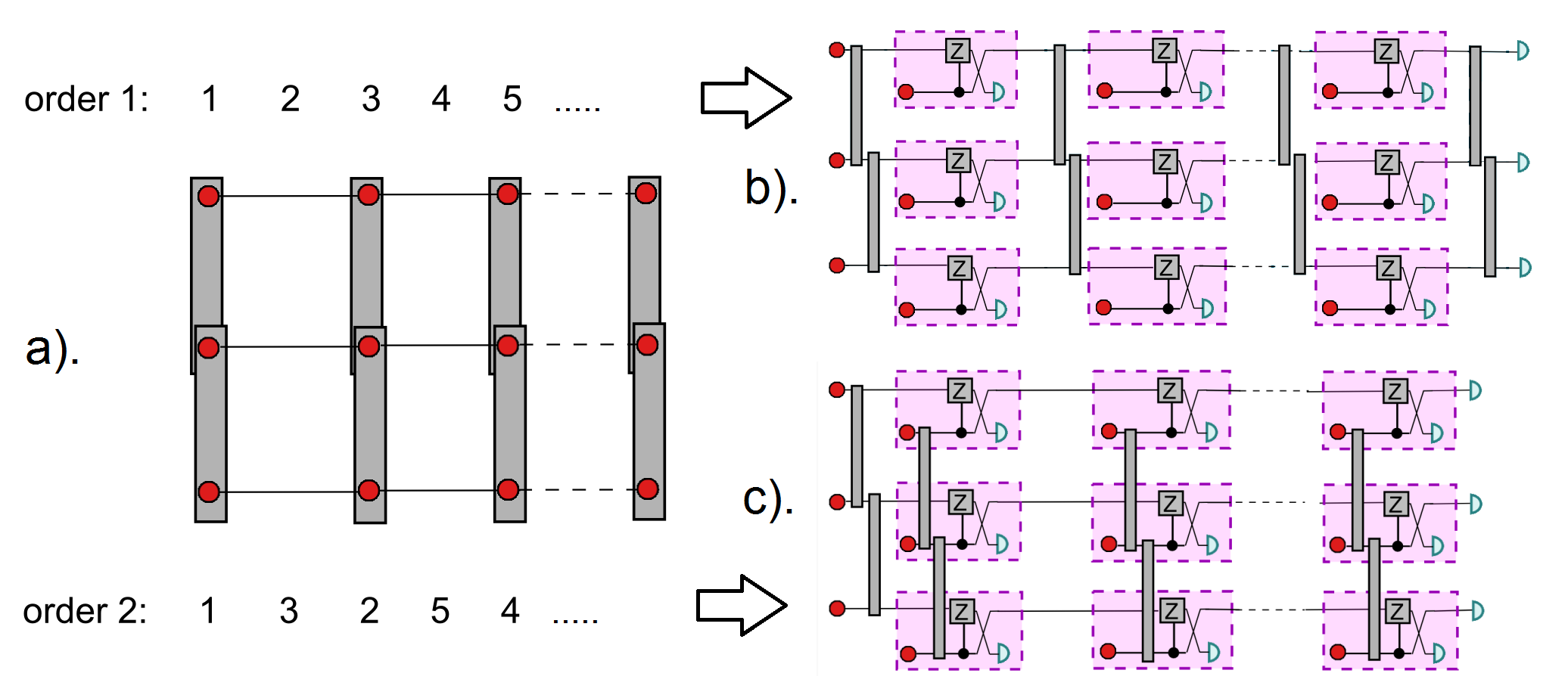}
\caption{2D examples of mappings from spatial to spatio-temporal correlations.
\textbf{a}) `Order 1' means that the state of the lattice is generated by applying the timelike (horizontal) gates before the respective next column of spacelike (vertical) gates. `Order 2' means that the spacelike gates are performed before the timelike gates connecting them.
\textbf{b}) Result of applying the map to `Order 1' case. The spatial correlations of measurement outcomes of local projective measurements performed on each node of the lattice are mapped into temporal correlations of outcomes of local POVM measurements performed on 1D lattice.
\textbf{c}) Result of applying the map to `Order 2' case.  The spatial correlations are  mapped into temporal correlations of outcomes of POVMs on 1D lattice, however in this case the POVM's have entangled ancillas. In both cases at the last stage projective measurements have to be performed.}
\label{SGSfig}
\end{figure*}

\subsection*{Mapping from spatial to spatio-temporal correlations for multidimensional lattices}

Although every finite-dimensional multipartite quantum state can be expressed as an MPS, the mapping from spatial to temporal correlations is not always possible. Whenever the bond dimension of an MPS exceeds the dimension of a single quantum system under consideration, multipartite unitaries are necessary for the sequential generation. Our mapping, however, works for bipartite unitaries only. To broaden the class of states for which the map can be applied we consider an $r$-dimensional hypercubic lattice, with a $d$-level quantum system at each node. We define two procedures  mapping all spatial correlations in such state to spatio-temporal ones (see Fig. 1). In both cases one of the spatial dimensions is distinguished, and the sequential generation is performed with respect to it. 
The initial state of the entire system is a product state.

In the first case (see Fig. 1a, order 1 of operations on the hypercube), at each stage $i$  we: 1)  generate an arbitrary quantum state of the particles placed in an $i$-th $(r-1)$-dimensional slice perpendicular to the distinguished \emph{temporal} dimension of the hypercube;
 2) move in temporal direction and perform all the unitaries between $i$-th and $i+1$-st slice that is next in time;
 3) perform projective measurements. 
 We consecutively repeat steps 1-3. 
All the measurements can be mapped into a single particle POVMs in complete analogy to the 1D case. 
The difference is that the initial input state for the local POVMs at each stage is the entire state of the particles of the $(r-1)$-dimensional hypercube.
An important example of a state with correlations that can be mapped in this way is a 2D cluster state \cite{BR01}, which is created by applying C-phase gates between every two neighbouring  qubits on a square lattice. 
Therefore a sequence of projective measurements on a 2D cluster state can be mapped into a sequence of POVMs performed on repeatedly created 1D cluster (Fig. 1b).

In the second procedure (Fig. 1a, order 2 of operations on the hypercube), one first creates arbitrary quantum states of the two neighbouring $(r-1)$-dimensional slices perpendicular to the distinguished dimension, then the unitaries inbetween them are performed. 
In this case the correlations of measurements performed along temporal direction can also be mapped into correlations arising from sequence of POVMs. 
The important difference is that in this case at each stage  one firstly has to prepare the state of the $(r-1)$-dimensional slice, and use its particles as ancillas of the POVMs (Fig. 1c). Note that in this case one need not perform any entangling operations on output states of the consecutive POVMs. 
An interesting example of this class of states are the so called Sequentially Generated States (SGS) \cite{BGWVC08}, which are a subclass of PEPS (Projected Entangled Pair States \cite{PEPS}.
PEPS provide a complete representation of arbitrary quantum state of finite dimension in terms of a 2D tensor network. SGS's are characterized by the property, that, in contrast to arbitrary PEPS, they have exponentially vanishing long-range correlations \cite{BGWVC08}.

Note that we generalize the standard notion of a graph state \cite{HEB04} to the case of non-commuting gates, which to the best of our knowledge was firstly suggested in \cite{GE07}. In the case the unitaries in the temporal direction commute with the unitaries needed to create the state of the $(r-1)$-dimensional slice, both mappings can be used.

\bibliographystyle{apsrev4-1}
\bibliography{ccp}

\end{document}